\documentclass[a4paper, amsfonts, amssymb, amsmath, reprint, showkeys, nofootinbib, twoside, superscriptaddress]{revtex4-1}
\usepackage[english]{babel}
\usepackage[utf8]{inputenc}
\usepackage[nodisplayskipstretch]{setspace}
\usepackage[colorinlistoftodos, color=green!40, prependcaption]{todonotes}
\usepackage{amsthm}
\usepackage{mathtools}
\usepackage{physics}
\usepackage{xcolor}
\usepackage{graphicx}
\usepackage[margin=0.75in]{geometry}
\usepackage{adjustbox}
\usepackage{placeins}
\usepackage[T1]{fontenc}
\usepackage{lipsum}
\usepackage{csquotes}
\usepackage{algorithm2e}
\usepackage[pdftex, pdftitle={Article}, pdfauthor={Author}]{hyperref} 
\begin{document}
\title{Kernel Alignment for Quantum Support Vector Machines Using Genetic Algorithms}

\author{Floyd M. Creevey}
\email{floyd.creevey@unimelb.edu.au}
\affiliation{School of Physics, University of Melbourne, VIC, Parkville, 3010, Australia.}
\author{Jamie A. Heredge}
\affiliation{School of Physics, University of Melbourne, VIC, Parkville, 3010, Australia.}
\author{Martin E. Sevior}
\affiliation{School of Physics, University of Melbourne, VIC, Parkville, 3010, Australia.}
\author{Lloyd C. L. Hollenberg}
\affiliation{School of Physics, University of Melbourne, VIC, Parkville, 3010, Australia.}

\date{\today} 

\begin{abstract}

    The data encoding circuits used in quantum support vector machine (QSVM) kernels play a crucial role in their classification accuracy. However, manually designing these circuits poses significant challenges in terms of time and performance. To address this, we leverage the GASP (Genetic Algorithm for State Preparation) framework for gate sequence selection in QSVM kernel circuits. We explore supervised and unsupervised kernel loss functions' impact on encoding circuit optimisation and evaluate them on diverse datasets for binary and multiple-class scenarios. Benchmarking against classical and quantum kernels reveals GASP-generated circuits matching or surpassing standard techniques. We also analyse the relationship between test accuracy and quantum kernel entropy, with results indicating little correlation. Our automated framework reduces trial and error, and enables improved QSVM based machine learning performance for finance, healthcare, and materials science applications.
    
\end{abstract}


\maketitle
\section{Introduction} \label{sec:Introduction}

Support vector machines (SVMs) are a popular class of classical machine learning algorithms widely used for classification and regression tasks in various fields such as finance \cite{kurani_comprehensive_2023}, healthcare \cite{yu_application_2010, venkatesan_ecg_2018}, and chemistry \cite{guangli_predicting_2006}. They work by identifying a hyperplane that best separates the given data. Recently, quantum SVMs (QSVMs) were introduced as a generalisation of SVMs that use a quantum kernel function to measure the similarity between data points \cite{havlicek_supervised_2019}. The quantum kernel function is typically implemented using a quantum circuit that encodes the features of the data points into the amplitudes of a quantum state, and then applies a set of quantum gates to compute the inner product between the quantum states. One advantage of QSVMs over classical SVMs is their ability in principle to construct kernels that are not classically simulable \cite{rebentrost_quantum_2014}. Another advantage is their potential for enhanced performance on certain tasks, such as the classification of non-linearly separable data, due to the ability of the quantum kernel function to operate in a higher-dimensional feature space than classical SVMs \cite{noble_what_2006}. These potential advantages make QSVMs an attractive area of research for improving the performance of classical machine learning algorithms. However, designing effective quantum kernel functions for given situations/data sets is a non-trivial task, and has been the subject of much research in the field of quantum machine learning \cite{altares-lopez_automatic_2021}. 

Several techniques are currently used to design quantum circuits for QSVM kernels, including analytical \cite{benedetti_parameterized_2019}, numerical \cite{suzuki_analysis_2020}, and variational methods \cite{wang_automated_2023, nakaji_approximate_2022}. Each of these techniques has its strengths and weaknesses, with the choice of technique depending on the specific problem and the available resources. Analytical methods are based on mathematical analysis and aim to determine the optimal kernel circuit design that minimises the error in the output. These methods can be computationally efficient, but they may not be suitable for all problems, as they rely on the availability of exact mathematical solutions. Numerical optimisation techniques involve the use of optimisation algorithms to find the best circuit design by minimising a cost function that measures the error in the output. These methods are generally more flexible than analytical methods, but they can be computationally intensive and may require large amounts of data. Variational methods employ parameterised quantum circuits, where the parameters are optimised to minimise output error, based on the idea of representing the quantum state through a neural network and optimising network parameters using classical algorithms. Such methods are highly flexible and can be used for a wide range of problems, but they may not always produce the most accurate results. 

The approach toward optimising the quantum circuit implementing the quantum kernel function described in this work is to use a genetic algorithm, building from the GASP framework presented in \cite{creevey_gasp_2023}. Previous work has been done on the generation of optimal ad hoc kernel function quantum circuits for classification using a QSVM  \cite{altares-lopez_automatic_2021, altares-lopez_autoqml_2022}, via the use of a genetic algorithm and comparison with classical classifiers. This work differs from previous work by examining different distinct fitness functions for the generation of the kernel function and analysing performance based on both their classification accuracy and entropy of entanglement, and utilising the gate set native to IBM hardware. The genetic algorithm can be used to search for an optimal quantum circuit that produces an accurate kernel matrix for the SVM while taking into account various constraints such as the number of qubits, the gate depth, and the connectivity of the quantum circuit. 

To generate the quantum circuits, a set of four gates: the single-qubit $X$, $\sqrt{X}$, and $R_z$ gates, and the two-qubit $CNOT$ gate are used. These form a universal gate set for quantum computation and are commonly used in quantum algorithms \cite{nielsen_quantum_2010}. The performance of the kernel function generated by the genetic algorithm using the accuracy of classification results on the testing set is evaluated and compared with classical kernels and quantum kernels. 

The results obtained here show that the kernel function quantum circuits generated by the genetic algorithm using this gate set perform comparably or better than classical kernels \cite{thurnhofer-hemsi_radial_2020}, and consistently outperform the standard PauliZZ quantum kernel \cite{schuld_quantum_2019}. This approach additionally aligns kernel circuits in a manner that enables them to traverse the Hilbert space region where the solution classification is situated. These results demonstrate that employing a genetic algorithm for quantum circuit optimisation offers a promising alternative to manually designing quantum circuits and can be extended to larger and more complex datasets.

The remainder of this paper will have the following structure. Section \ref{sec:background} will give a summary of SVMs, QSVMs and genetic algorithms. Section \ref{sec:methods} will describe the proposed method for kernel generation in detail. Section \ref{sec:results} will present the results, and section \ref{sec:conclusion} will present the conclusions and potential future work.

\section{Quantum Classification} \label{sec:background}

  Here we outline the key concepts required for the work. Quantum machine learning relies heavily on classical machine learning. As such a background of both the classical, and quantum, machine learning methods will be outlined.

  \subsection{Support Vector Machines}
  
\begin{figure}
    \centering
    \includegraphics[width=0.5\textwidth]{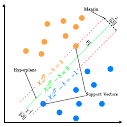}
    \caption[SVM Diagram]{Basic overview of a binary classification SVM. The two classes are represented by orange and blue circles, respectively. The hyperplanes for $1$, $0$, and $-1$ are represented by orange, green, and blue lines, respectively. The margin is the distance between support vectors of the two classes, $\frac{2}{||\vec{\omega}||}$. The distance to each statevector from the hyperplane is $\frac{b}{||\vec{\omega}||}$.}
    \label{fig:svm_diagram}
  \end{figure}

  SVMs \cite{flach_machine_2012}  are a popular classical supervised machine learning method for classification and regression tasks. SVMs aim to find a hyperplane that separates the data into two or more classes, with the largest possible margin between the closest data points to the hyperplane. These closest data points are called support vectors, and the margin is defined as the perpendicular distance between the hyperplane and the closest support vectors. The intuition behind this approach is that the larger the margin, the more robust the classifier will be to new data. Consider a dataset $\mathcal{D}$ of $n$ data points each with $m$ features, of the form $(\mathcal{X}, \vec{y})$, where,
  \begin{equation}
      \mathcal{X} = \begin{pmatrix}
      \vec{X}_1 \\
      \vec{X}_2 \\
      \vdots \\
      \vec{X}_n \\
  \end{pmatrix} = \begin{pmatrix}
      x_{1, 1} & x_{1, 2} & \hdots & x_{1, m} \\
      x_{2, 1} & x_{2, 2} & \hdots & x_{2, m} \\
      \vdots & \vdots & \ddots & \vdots \\
      x_{n, 1} & x_{n, 2} & \hdots & x_{n, m} \\
  \end{pmatrix},
  \end{equation}
  where $\vec{X}_i$ is a datapoint in $\mathcal{X}$, and $x_{i, j}$ is a feature of $\vec{X}_i$, and,
  \begin{equation}
      \vec{y} = \begin{pmatrix}
      y_1 \\
      y_2 \\
      \vdots \\
      y_n \\
  \end{pmatrix},
  \end{equation}
  where $y_i$ is the data label for $\vec{X}_i$ in $\vec{y}$. Any hyperplane can be written with $\mathcal{X}$ satisfying,
  \begin{equation}
    \mathcal{X}\vec{\omega}^{T}-b=0,
  \end{equation}
  where $\vec{\omega}$ are the weights that form a normal vector to the hyperplane, and $b$ is the bias used to determine the offset of the hyperplane from the origin along $\vec{\omega}$, $\frac{b}{||\vec{\omega}||}$. For linearly separable two-class data, two hyperplanes can be determined to maximise the distance between the classes, described by,
  \begin{equation}
    \mathcal{X}\vec{\omega}^{T}-b=1,
  \end{equation}
  with anything on or above this boundary being of one class, and,
  \begin{equation}
    \mathcal{X}\vec{\omega}^{T}-b=-1,
  \end{equation}
  with anything on or below this boundary being of the other class. The region between these hyperplanes is the margin, and the maximum margin hyperplane, often simply the hyperplane, is halfway between them. To determine the margin width, let $\vec{\mathcal{X}}_+$ be the closest point to the hyperplane of class $+$, and $\vec{\mathcal{X}}_-$ be the closest point to the hyperplane of class $-$, so the width would be,
  \begin{equation}
    (\vec{\mathcal{X}}_+ - \vec{\mathcal{X}}_-)\cdot\frac{\vec{\omega}}{||\vec{\omega}||} = \frac{1 - b + 1 + b}{||\vec{\omega}||} = \frac{2}{||\vec{\omega}||}.
  \end{equation}
  A diagram displaying these concepts can be seen in Figure \ref{fig:svm_diagram}.

  The first step  in solving the SVM is to  define the SVM Lagrangian, also called the primal problem,
  \begin{equation}\label{eqn:primal}
      \begin{aligned}
          \mathcal{L}(\vec{\omega}, b, \vec{\alpha}) = \frac{1}{2}\vec{\omega}^T\cdot\vec{\omega} &- \sum_{i=1}^n\alpha_iy_i\vec{X}_i^T\cdot\vec{\omega}\\ &+ b\sum_{i=1}^n\alpha_iy_i + \sum_{i=1}^n\alpha_i,
      \end{aligned}
  \end{equation}
  where $\vec{\omega}$ are the weights, $b$ is the bias, and $\vec{\alpha}$ are the Lagrange multipliers. Then compute the partial derivatives with respect to its primal variables, $\vec{\omega}$ and $b$,
  \begin{equation}
      \begin{aligned}
          \frac{\partial\mathcal{L}}{\partial \vec{\omega}} = 0 \rightarrow \vec{\omega}^* &= \sum_{i=1}^n\alpha_iy_i\vec{X}_i,
      \end{aligned}
  \end{equation}
  where $^*$ is the complex conjugate, and,
  \begin{equation}
      \begin{aligned}
          \frac{\partial\mathcal{L}}{\partial b} = 0 \rightarrow 0 &= \sum_{i=1}^n\alpha_iy_i.
      \end{aligned}
  \end{equation}
  With this, the Lagrangian dual of the primal problem, Equation \ref{eqn:primal}, 
  \begin{equation}
      \begin{aligned}
          \alpha_1^*, \alpha_2^*, \ldots, \alpha_n^* = \max_{\alpha_1, \alpha_2, \ldots, \alpha_n}\mathcal{L}(\vec{\omega}^*, b^*, \vec{\alpha}),
      \end{aligned}
  \end{equation}
  can be solved as,
  \begin{equation}
    \begin{aligned}
        \mathcal{L}(\vec{\omega}^*, b^*, \vec{\alpha}) = -\frac{1}{2}\sum_{i=1}^n\sum_{j=1}^n\alpha_i\alpha_jy_iy_j\vec{X}_i^T\cdot\vec{X}_j + \sum_{i=1}^n\alpha_i,
    \end{aligned}
\end{equation}
  subject to $\alpha_i \ge 0$, $i \in 1, 2, \ldots, n$ and $\sum_{i=1}^n\alpha_iy_i=0$. This allows the  Lagrange multipliers,  $\alpha_i$,  can be found, and used to find the weights  $\vec{\omega}^*$,
  \begin{equation}
      \begin{aligned}
          \vec{\omega}^* &= \sum_{i=1}^n\alpha_iy_i\vec{X}_i.
      \end{aligned}
  \end{equation}
  For linear multi-class classification, the obvious strategy might be to create a two-class classifier that distinguishes points in a particular class $c$ from all other classes, i.e points in class $c$ are labeled as $(+1)$ and points not in class $c$ as $(-1)$. Then solve for all $C$ classes, 
  \begin{equation}
      \begin{aligned}
          \vec{X}^T\cdot\vec{\omega}_c^* + b_c^* = 0
      \end{aligned}
  \end{equation}
  with $c = 1, 2, \ldots, C,$ so that all points from class $c$ will lie on the positive side of its decision boundary $(w_c^*, b_c^*)$, while points from other classes lie on its negative side. Hence, datapoint $\vec{X}_i$ belongs to class $c$ if it satisfies the two following inequalities,
  \begin{equation}
      \begin{aligned}
          &\vec{X}_i^T\cdot\vec{\omega}_c^* + b_c^* > 0 \\
          &\vec{X}_i^T\cdot w_j^* + b_j^* < 0 \text{ for all } j\ne c.
      \end{aligned}
  \end{equation}
  However,  this is typically not a good approach, as it does not allow points in ambiguous space between bounds to be assigned labels \cite{pal_multiclass_2008}. A solution to this is to use the fusion rule. The fusion rule generalises to assign a label for each point  $\vec{X}_i$ by finding not the classifier that produces a positive evaluation $\vec{X}_i^T\cdot\vec{\omega}_c^*+b_c^* > 0$, but by assigning $\vec{X}_i$ the class label $c$ with the largest evaluation (even when negative),
  \begin{equation}
      \begin{aligned}
          y = \max_{c=1, 2, \ldots, C}\vec{X}_i^T\cdot w_c^* + b_c^*.
      \end{aligned}
  \end{equation}
  This assigns labels to the entire space and effectively handles overlapping classes. Using the fusion rule, $C$ individual classifiers are learned, each distinguishing one class from the remainder of the data. The learned classifiers are then combined to make final assignments.
  
  For a non-linear classification, first a good feature transformation $\phi$ must be found to map the data points into linearly separable sets,
  \begin{equation}
      \begin{aligned}
          \mathcal{L}(\vec{\omega}^*, b^*, \vec{\alpha}) &= \sum_{i=1}^n\alpha_i - \frac{1}{2}\sum_{i=1}^n\sum_{j=1}^n\alpha_i\alpha_jy_iy_j\phi(\vec{X}_i)^T\phi(\vec{X}_j),
      \end{aligned}
  \end{equation}
   subject to $\alpha_i \ge 0$, $i \in 1, 2, \ldots, n$ and $\sum_{i=1}^n\alpha_iy_i=0$, where $\phi(\vec{X}_i)^T\phi(\vec{X}_j)$ is a similarity measure between the transformed datapoints. The kernel function $\mathcal{K}(\vec{X}_i, \vec{X}_j) = \phi(\vec{X}_i)^T\phi(\vec{X}_j)$ can then be identified, which defines the inner products in the transformed space. This is known as the kernel trick.

  \subsection{Quantum Support Vector Machines}
  
\begin{figure*}
    \centering
    \includegraphics[width=0.9\textwidth]{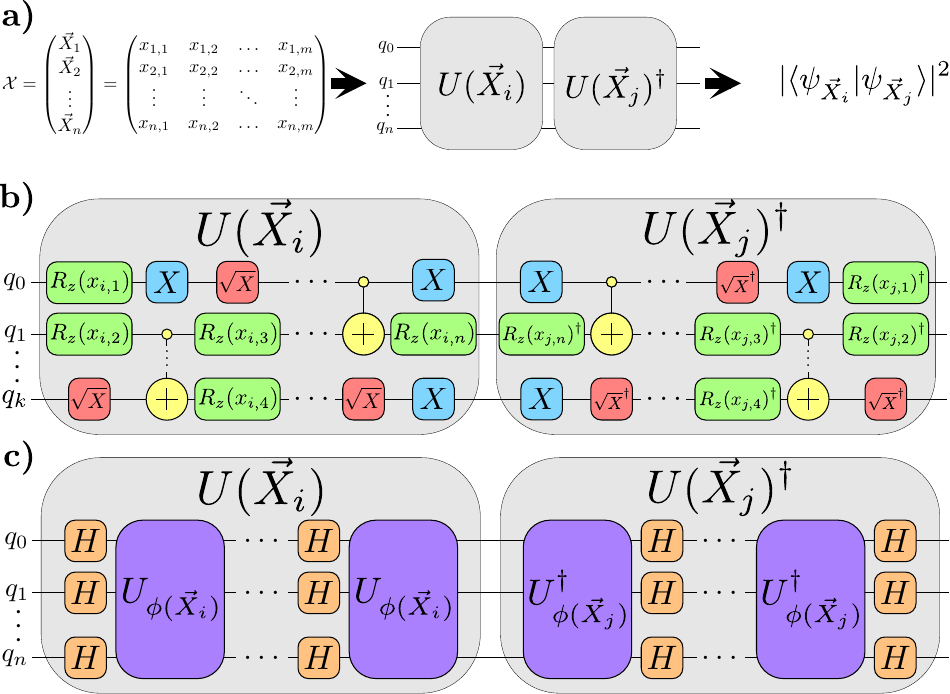}
    \caption[Basic Overview of QSVMs]{Basic overview of QSVMs. \textbf{a)} The datapoints, $\mathcal{X}$, with labels $\vec{y}$, are used to construct the elements of the kernel matrix, defined as the piecewise inner product, for the QSVM using the quantum kernel function for the given circuit structure. \textbf{b)} An example circuit that could be produced by the GA method. The GA is stochastic by nature, the number of features used in the circuit will vary, as the circuit structure will likely be different each time a circuit is generated. In the GA circuit, the $X$ and $\sqrt{X}$ gates represent the Pauli $X$ and the square root of the Pauli $X$ respectively, not to be confused with the input data $\vec{X_i}$. \textbf{c)} Circuit structure of the PauliZZ quantum kernel function where $H$ symbolises the Hadamard gate, and $U_{\phi(\vec{X}_i)} = \exp(i\sum_{S\in I}\phi_S(\vec{X}_i)\prod_{i\in S}P_i)$.}
    \label{fig:QSVM_overview}
  \end{figure*}

  QSVMs are a quantum computing-based variant of SVMs. QSVMs leverage the principles of quantum computing to potentially offer advantages in handling high-dimensional data and solving complex optimisation problems. Consider the same dataset $\mathcal{D}$ as above. In QSVMs, input data points are encoded as quantum states. These quantum states are represented as  $|\psi_{\vec{X}}\rangle$. Each data point $\vec{X}$ is mapped to a quantum state $|\psi_{\vec{X}}\rangle$ as,
  \begin{equation}
      |\psi_{\vec{X}}\rangle = U(\vec{X})|0\rangle.
  \end{equation}
  The central element of QSVMs is the quantum kernel function, which quantifies the similarity between quantum states. The quantum kernel function is defined as,
  \begin{equation}
      \mathcal{K}(\vec{X}_i, \vec{X}_j) = |\langle\psi_{\vec{X}_i}|\psi_{\vec{X}_j}\rangle|^2,
  \end{equation}
  i.e. the inner product of the quantum states $|\psi_{\vec{X}_i}\rangle$ and $|\psi_{\vec{X}_j}\rangle$ (see Figure \ref{fig:QSVM_overview}a). This is analogous to the kernel trick in classical SVMs but operates in a quantum state space.

  The construction of the quantum kernel is dependent on the choice of $U$. A commonly used quantum kernel is repeated layers of the Pauli expansion circuit \cite{havlicek_supervised_2019}. It can be constructed as,
  \begin{equation}
      \begin{aligned}
          H^{\otimes n}U_{\phi(\vec{X}_i)},
          \end{aligned}
  \end{equation}
  where,
  \begin{equation}
    U_{\phi(\vec{X}_i)} = \exp(i\sum_{S\in I}\phi_S(\vec{X}_i)\prod_{i\in S}P_i),
  \end{equation}
  $S$ is a set of qubit indices describing the connections in the feature map, $I$ is a set containing all index sets, $P_i\in\{I, X, Y, Z\}$, and the data mapping, $\phi_S$ is,
  \begin{equation}
    \phi_S(\vec{X}_i)=\begin{cases}
      \vec{X}_i \ \text{if $S=\{i\}$}, \\
      \prod_{j\in S}(\pi-\vec{X}_j) \ \text{if $|S|>1$.}
    \end{cases}
  \end{equation}
  A basic overview of QSVMs is displayed in Figure \ref{fig:QSVM_overview}.

  \subsection{Quantum Neural Networks}

  A Quantum neural network (QNN) is a parameterised quantum circuit (PQC), generally comprised of a feature map and an ansatz, or variational circuit layer. A PQC is defined by a sequence of quantum gates, some of which have adjustable parameters. A feature map encodes the data, or input parameters, into the quantum computer. An ansatz contains trainable weights to adjust the hyperplane and separate the data in this encoded space. The goal is to find the optimal set of parameters, $\vec{\theta}$, that minimises a given cost function. Mathematically, a PQC can be represented as,
  \begin{equation}
    U(\vec{\theta}) = U_n(\theta_n)U_{n-1}(\theta_{n-1})\ldots U_1(\theta_1),
  \end{equation}
  where $n$ is the total number of parameterised gates in the circuit.

\begin{figure}[h]
    \centering
    \includegraphics[width=0.5\textwidth]{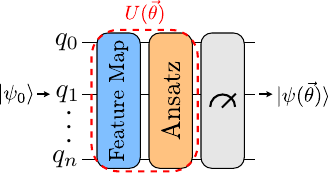}
    \caption[Basic Overview of QNNs]{Basic overview of QNNs. QNNs are composed of a feature map and an ansatz. The feature map encodes data into the quantum circuit, while the ansatz is a variational layer that adjusts the hyperplane between classes of data to find the optimal classification.}
    \label{fig:qnn_overview}
  \end{figure}

  The initial state of the QNN, $|\psi_0\rangle$, is transformed through the PQC to the result in the final state, $|\psi(\vec{\theta})\rangle$, given by,
  \begin{equation}
    |\psi(\vec{\theta})\rangle = U(\vec{\theta})|\psi_0\rangle.
  \end{equation}
  This final state encodes the outcome of the quantum system, influenced by the parameterised gates. This structure is highlighted in Figure \ref{fig:qnn_overview}.

  The performance of the QNN is evaluated using a cost function, $C(\vec{\theta})$, which depends on the output state $|\psi(\vec{\theta})\rangle$ and is designed according to the specific problem being solved. In  many quantum algorithms, particularly in quantum simulations and optimisation problems, the cost function is  the expectation value (or energy)  of a problem-specific Hamiltonian, $H$, with respect to the parameterised state,
  \begin{equation}
    C(\vec{\theta}) = \langle\psi(\vec{\theta})|H|\psi(\vec{\theta})\rangle.
  \end{equation}

  In classification or regression, the cost function might measure the distance between the quantum state produced by the QNN and a target state or classical data representation. Examples include the fidelity between the output state and a desired target state,
  \begin{equation}
    C(\vec{\theta}) = |\langle\psi_\text{target}|\psi(\vec{\theta})\rangle|^2,
  \end{equation}
  or more classical metrics such as mean squared error (MSE), when the output can be directly related to classical labels. For classification tasks, the cost function can be based on the probabilities of measurement outcomes. For instance, if the QNN is designed to classify inputs into two categories, the cost function might be formulated to maximise the probability of correctly classifying input points, using metrics such as cross-entropy loss.

  The optimisation of $\vec{\theta}$ aims to find the parameter values that minimise the cost function $C(\vec{\theta})$. This is typically achieved through classical gradient-based optimisation techniques. The gradient of the cost function with respect to the parameters, $\nabla_{\vec{\theta}} C(\vec{\theta})$, guides the update of $\vec{\theta}$ across iterations,
  \begin{equation}
    \vec{\theta}_{\text{new}} = \vec{\theta}_{\text{old}}-\eta\nabla_{\vec{\theta}} C(\vec{\theta}),
  \end{equation}
  where $\eta$ is the learning rate, a hyperparameter that controls the size of the steps taken during optimisation. For certain PQCs, it is possible to compute the gradient using quantum techniques, such as the parameter shift rule,
  \begin{equation}
    \frac{d\langle H\rangle}{d\vec{\theta}} = \frac{1}{2}\left(\langle H\rangle_{\vec{\theta} + \frac{\pi}{2}} - \langle H\rangle_{\vec{\theta} - \frac{\pi}{2}}\right),
  \end{equation}
  where $\langle H\rangle_{\vec{\theta}}$ is the expectation of $H$ when parameterised by $\vec{\theta}$. This approach can be more efficient than classical finite difference methods for gradient estimation in quantum systems.

  \subsection{Definitions}

  Finally, we note some important definitions.
  
  \textbf{\textit{Eigenvalue Analysis of Kernel Matrices}} assesses the fitness of a kernel matrix, $\mathcal{K}$, created with the corresponding kernel circuit, and optimises by maximising its largest normalised eigenvalue. Let $\lambda_1, \lambda_2, \ldots, \lambda_n$ be the eigenvalues of $\mathcal{K}$, where $n$ is the dimension of $\mathcal{K}$, the maximum normalised eigenvalue of is $\frac{\lambda_1}{n}$. This method is motivated by Spectral Clustering \cite{ng_spectral_2001}, and is similar to kernel principal component analysis (KPCA) \cite{scholkopf_kernel_1997}. Maximising the largest normalised eigenvalue can lead to a focus on the dominant mode in the data, potentially resulting in dimensionality reduction. 

  \textbf{\textit{Entropy of Entanglement}} of a kernel circuit is calculated with the Von Neumann Entropy as,
  \begin{equation}
      S = \frac{\sum_{i=1}^n-\rm{Tr}(\rho_i\ln\rho_i)}{n}
  \end{equation}
  \noindent where $n$ is the number of qubits in the circuit, and $\rho_i$ is the partial trace of the kernel matrix with respect to the $i$th qubit.

\section{Implementation of QML Algorithms using GASP}\label{sec:methods}

This section will describe how genetic algorithms were used to generate encoding circuits for QSVMs and feature maps for QNNs. The framework used is an extension to GASP, highlighting the versatility of its use in various aspects of quantum computing, which has been termed `Genetically Engineered Kernel Optimisation', or `GEKO'.

\subsection{Quantum Support Vector Machines}

  For QSVMs, we first perform classification on several synthetic datasets, moons, XOR, and circles, before performing classification on real datasets, Iris, Wine, and Breast Cancer, all from scikitlearn \cite{pedregosa_scikit-learn_2011}. Finally, classification on the more challenging real datasets of Irrigation \cite{patel_h_intelligent_2020}, Drug Classification \cite{tripathi_p_drug_2020}, and Parkinson's \cite{little_exploiting_2007} is performed. 

  The datasets are all initially analysed with principal component analysis (PCA) \cite{pearson_liii_1901} to determine the number of components required to express $95\%$ of the variance in the data, and then used to reduce the number of components to that size. With this processed data, the method then uses GEKO, with each circuit having 5 qubits, to optimise the kernel circuit for the QSVM. Starting with an initial individual, which is a quantum circuit composed of gates from the gate set $\mathcal{G} = {X, \sqrt{X}, R_z, CNOT}$, the genetic algorithm then creates $k$ mutated individuals by creating $k$ copies of the individual, then mutating each individual with a probability $m_p$, set to $50\%$ for these results. For the supervised GEKO technique, each individual's fitness is assessed as the classification accuracy of running the QSVM with the corresponding kernel circuit on the training data. For the unsupervised GEKO technique, each individual's fitness is assessed by eigenvalue analysis of the kernel matrix.
  
  It is important to note here that the terms `supervised' and `unsupervised' refer to the technique for optimisation of the kernel circuit by genetic algorithm. The `supervised' technique is supervised as it requires data labels in the assessment of the kernel circuit classification accuracy for its fitness. The `unsupervised' technique does not use labels in the fitness assessment of the kernel circuit; the fitness is assessed with only the kernel matrix, as the maximisation of the largest normalised eigenvalue. However, the overall methodology of training support vector machines, whether they are classical or quantum, is always a supervised task.

  \begin{figure*}       
    \centering
    \includegraphics[width=\textwidth]{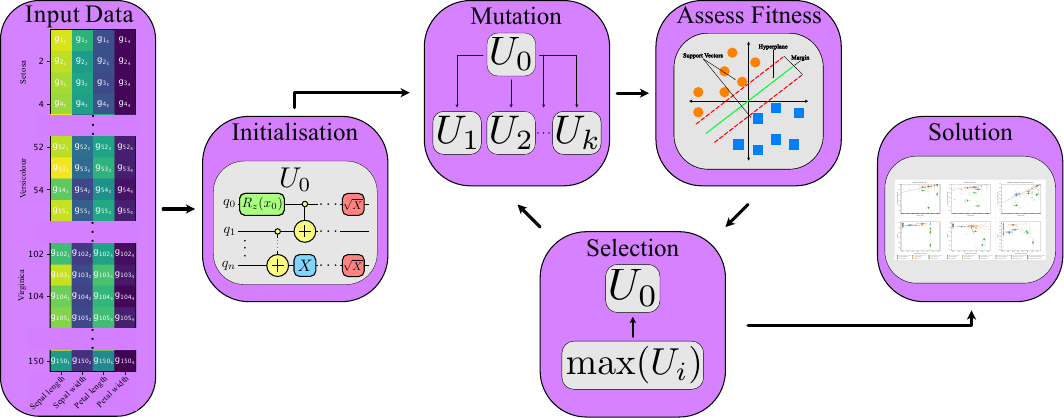}
    \caption[Overview of the GEKO for QSVM Approach]{Overview of the GEKO for QSVM approach, given a dataset. First, create an initial individual $U_0$ which is a quantum kernel circuit. Then mutate the individual with probability $p=50\%$ $k$ times, to produce a $k$ size population of individuals. Then assess the fitness of the quantum kernel circuits determined by each individual in the population: if supervised, the fitness is the result of training a QSVM using the given kernel, if unsupervised, the fitness is the maximum normalised eigenvalue of the kernel matrix of the circuit. Then select the most fit individual in the population, if it is more fit than $U_0$ it becomes $U_0$. Repeat until the desired fitness is achieved.}
    \label{fig:QSVM_flowchart}
  \end{figure*}
  
  The individual with the highest training fitness is then selected and their validation fitness is assessed in the same manner the training fitness was assessed, except validation data is used instead of training data. If the validation fitness of the most fit mutated individual is greater than the validation fitness of the current individual, it becomes the new current individual. This process is repeated for a fixed number of iterations, where each iteration is called a `generation', or until the individual's fitness reaches a desired threshold. If the desired threshold fitness is not achieved within the number of generations given, the number of genes is increased or decreased according to a binary search. This ensures that the individuals produced have the lowest amount of genes required to achieve the desired fitness.
  
  A flow chart of this methodology can be seen in Figure \ref{fig:QSVM_flowchart}, and pseudo-code for this methodology is shown in Algorithm \ref{alg:qsvm_asexual_genetic_algorithm}. Overall, the methods optimise the kernel function for the QSVM using a genetic algorithm with either a supervised or an unsupervised technique. The effectiveness of this method on nine benchmark datasets is demonstrated.

  \RestyleAlgo{ruled}
  \SetKwComment{Comment}{/* }{ */}
  \begin{algorithm}
  \scriptsize
      \caption{Genetic Engineered Kernel Optimisation (GEKO)}\label{alg:qsvm_asexual_genetic_algorithm}
      \KwData{$data$, $maximumGenerations$, $targetFitness$, $numberOfQubits$, $numberOfGenes$, $populationSize$, $optimisationTechnique$}
      \KwResult{$individual\quad(quantumCircuit)$}
      $minimum \gets 0$\;
      $maximum \gets \inf$\;
      $trainingData \gets \textbf{TrainingDataSplit(}data\textbf{)}$\;
      $validationData \gets \textbf{ValidationDataSplit(}data\textbf{)}$\;
      \While{$geneRange \ge numberOfQubits$}{
        $generations \gets 0$\;
        $bestFitness \gets 0$\;
        $bestIndividual \gets \textbf{Individual(}numberOfGenes\textbf{)}$\;
        $population \gets \{\quad\}\times populationSize$\;
        \While{$bestFitness < targetFitness\And generations \le maximumGenetations$}{
            \For{$i \gets 0$ \KwTo $populationSize$}{
                $newIndividual \gets \textbf{Mutate(}bestIndividual, 50\%\textbf{)}$\;
                \eIf{$optimisationTechnique ==\rm{ `Supervised'}$}{
                    $fitness \gets \textbf{QSVM(}newIndividual, trainingData\textbf{)}$\;
                }{$fitness \gets \textbf{MaximumEigenvalue(}newIndividual, trainingData\textbf{)}$\;}
                
                $population[fitness] = newIndividual$\;
            }
            $individual \gets \textbf{MaximumFitness(}population\textbf{)}$\;
            \eIf{$optimisationTechnique ==\rm{ `Supervised'}$}{
                $fitness \gets \textbf{QSVM(}individual, validationData\textbf{)}$\;
            }{$fitness \gets \textbf{MaximumEigenvalue(}individual, validationData\textbf{)}$\;}
            \eIf{$fitness \ge bestFitness$}{
                $bestFitness \gets fitness$\;
                $bestIndividual \gets individual$\;
                $maximum \gets numberOfGenes$\;
                $generations \gets 0$\;
            }{$minimum \gets numberOfGenes$;}
            $generations \gets generations + 1$\;
        }
        $geneRange = \textbf{BinarySearch(}minimum, maximum\textbf{)}$\;
      }
      \textbf{Return} $ind$\;
  \end{algorithm}

\subsection{Quantum Neural Networks}

\begin{figure*}
    \centering
    \includegraphics[width=\textwidth]{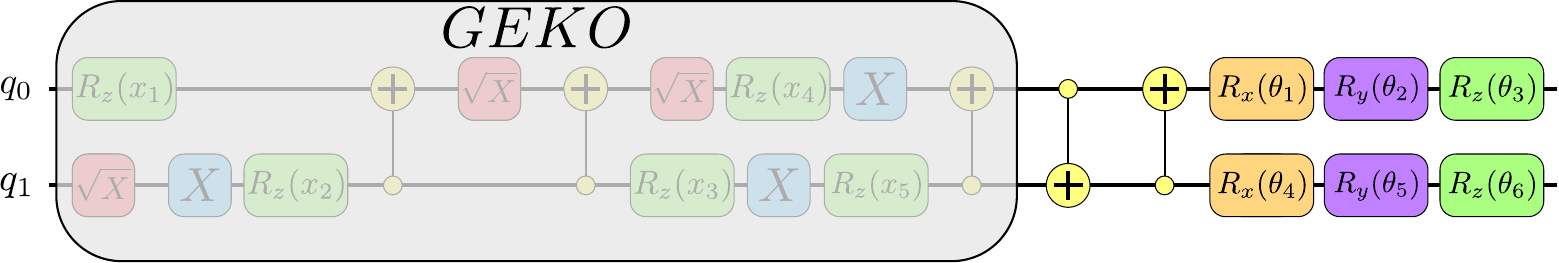}
    \caption[Example QNN Circuit]{Example QNN circuit. The circuit shown is comprised of a feature map and a variational layer. The feature map is generated with a genetic algorithm similar to GEKO, with an example of what could be the feature map shown faintly in the GEKO unitary, and the variational layer is comprised of entangling $CNOT$s and a series of $R_x(\theta_i)R_y(\theta_{i+1})R_z(\theta_{i+2})$ for each qubit.}
    \label{fig:QNN_circ}
  \end{figure*}

  For QNNs, classification was only conducted on the synthetic datasets, moons, XOR, and circles, from scikitlearn \cite{pedregosa_scikit-learn_2011}. Principal component analysis (PCA) \cite{pearson_liii_1901} was used to determine the number of components required to express $95\%$ of the variance in the data, and for the datasets used no reduction of the number of components was required. A  genetic algorithm was used to optimise the feature map for the QNN, with each circuit having $2$ qubits, as this was the amount needed to capture the maximum number of classes in any of the datasets used. Starting with an initial individual, which is a quantum circuit composed of gates from the gate set $\mathcal{G} = {X, \sqrt{X}, R_z, CNOT}$, the genetic algorithm then creates $k$ mutated individuals by creating $k$ copies of the individual, then mutating each individual with a probability $m_p$, set to $50\%$ for these results. After the feature map is generated, a variational circuit layer, comprised of linearly entangled $CNOT$ gates and a series of $R_x(\theta_i)R_y(\theta_{i+1})R_z(\theta_{i+2})$ on each qubit, was applied. An example of the resultant circuit is shown in Figure \ref{fig:QNN_circ}.

  Classical optimisation is then performed over this variational circuit layer, to adjust the hyperplane and identify the optimal classification of the data for the given feature map. Each individual's training fitness, $f$, is assessed as the error between the predicted data labels, $\hat{Y}$, and the true data labels, $Y$,
  \begin{equation}
    f = \frac{1}{n}\sum_i^n\begin{cases}
      \hat{Y}_i & \text{if}\ \hat{Y}_i=Y_i \\
      0 & \text{if}\ \hat{Y}_i\ne Y_i
    \end{cases}
  \end{equation}
  The individual with the highest training fitness is then selected and their validation fitness is assessed in the same manner as the training fitness, except validation data is used instead of training data. If the validation fitness of the most fit mutated individual is greater than the validation fitness of the current individual, it becomes the new current individual. This process is repeated for a fixed number of iterations, where each iteration is called a `generation', or until the individual's fitness reaches a desired threshold. If the desired threshold fitness is not achieved within the number of generations given, the number of genes is increased or decreased according to a binary search. This ensures that the individuals produced have the lowest number of genes required to achieve the desired fitness. Overall, the method optimises the feature map for the QNN using a genetic algorithm. The effectiveness of this method on three benchmark datasets is demonstrated.

\section{Generated Quantum Encoding Circuits}\label{sec:results}

  \subsection{Quantum Support Vector Machines}

\begin{figure}
    \centering
    \includegraphics[width=0.5\textwidth]{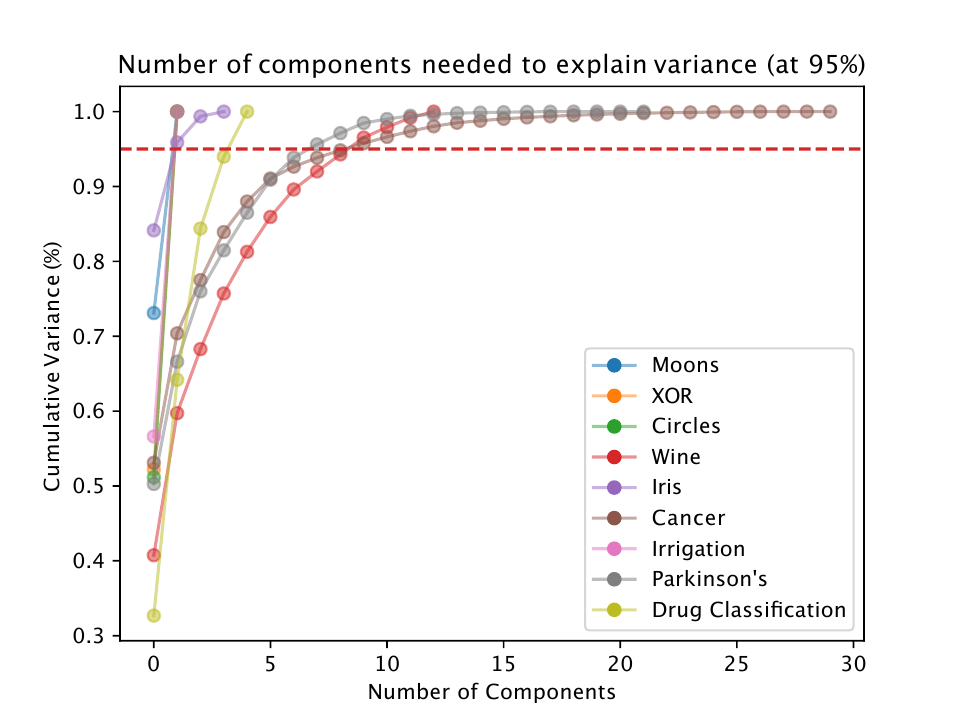}
    \caption[PCA Component Comparison]{Comparison of the number of features required to explain 95\% variance in the dataset for Moons, XOR, Circles, Wine, Iris, Cancer, Irrigation, Parkinson's, and Drug Classification datasets, coloured blue, orange, green, red, purple, brown, pink, grey, and light green, respectively. The dots represent features, ordered from the feature that explains the highest amount of variance in the data, to the feature that explains the lowest amount of variance in the data, the red dashed line represents 95\% of the data being explained.}
    \label{fig:pca_components_fig}
  \end{figure}

  The methods were tested on the Moons, XOR, Circles, Wine, Iris, Cancer, Irrigation, Parkinson's, and Drug Classification data sets. The Moons, XOR, and circles datasets were synthetically generated with 400 points each. The Wine, Iris, Cancer, Irrigation, Parkinson's and Drug Classification datasets were all real data, and all data points were used. PCA was used to determine the number of features required for each dataset to explain $95\%$ of the variance in the data. The results of this analysis can be seen in Figure \ref{fig:pca_components_fig}, and the number of features used for each dataset can be seen in Table \ref{tab:n_components_table}. 
  
  \begin{table*}
    \small
    \centering
    \caption{\label{tab:n_components_table}Number of components needed to explain 95\% variance in each dataset from analysis with PCA.}
    \begin{tabular}{@{}lll}
      \hline
      \hline
      Dataset & Total Number of Components & $95\%$ Variance Number of Components \\
      \hline
      Moons & 2 & 2 \\
      XOR & 2 & 2 \\
      Circles & 2 & 2 \\
      Wine & 13 & 10 \\
      Iris & 4 & 2 \\
      Cancer & 30 & 10 \\
      Irrigation & 2 & 2 \\
      Parkinson's & 22 & 8 \\
      Drug Classification & 5 & 5 \\
      \hline
      \hline
    \end{tabular}
  \end{table*}
  
  Each data set was split into $20\%$ testing data, $48\%$ training data, and $32\%$ validation data. The total number of data points and split of data for each dataset is displayed in Table \ref{tab:datasetsamples}.

  \begin{table*}
    \small
    \caption{\label{tab:datasetsamples}Dataset splits.}
    \begin{tabular}{@{}lllll}
      \hline
      \hline
      Dataset & Total points  & Training points  & Validation points  & Testing points  \\
      \hline
      Moons & 400 & 192 & 128 & 80 \\
      XOR & 400 & 192 & 128 & 80 \\
      Circles & 400 & 192 & 128 & 80 \\
      Wine & 178 & 85 & 57 & 36 \\
      Iris & 150 & 72 & 48 & 30 \\
      Cancer & 556 & 273 & 182 & 114 \\
      Irrigation & 200 & 96 & 64 & 40 \\
      Parkinson's & 195 & 93 & 63 & 39 \\
      Drug Classification & 200 & 96 & 64 & 40 \\
      \hline
      \hline
    \end{tabular}
  \end{table*}

\begin{figure*}
    \centering
    \includegraphics[width=\textwidth]{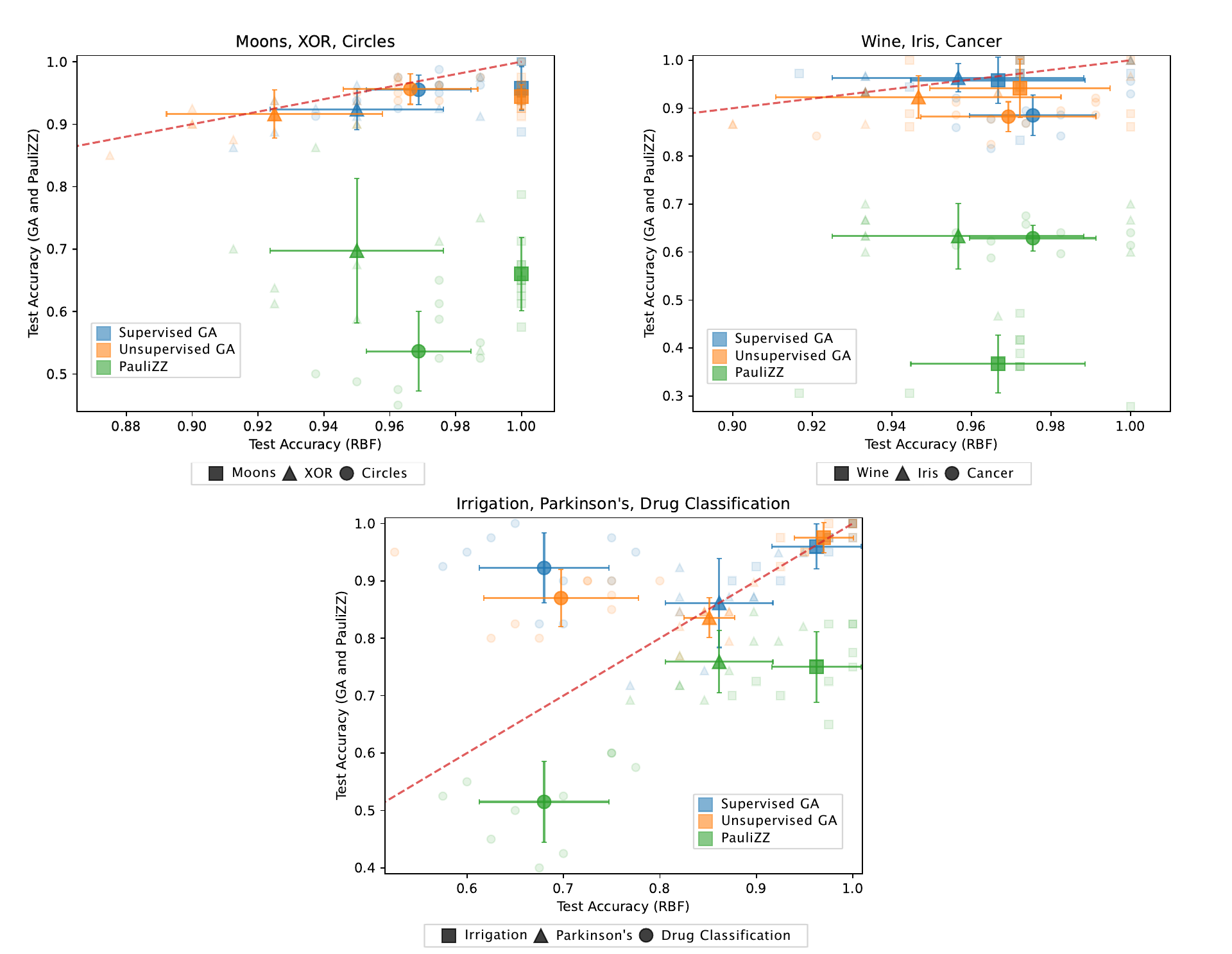}
    \caption[Comparison of Supervised GEKO, Unsupervised GEKO, PauliZZ, and RBF Generated Kernels]{Comparison of supervised GEKO, unsupervised GEKO, PauliZZ, and RBF generated kernels on Moons, XOR, Circles, Wine, Iris, Cancer, Irrigation, Parkinson's, and Drug Classification datasets. The supervised GEKO, unsupervised GEKO, PauliZZ, and RBF kernels are blue, orange, green, and grey, respectively. The plots show the comparison of the test accuracies achieved for the supervised GEKO, unsupervised GEKO, and PauliZZ kernels against the test accuracy achieved by the RBF kernel. The larger, bolder symbols for each dataset represent the average result over the 10 tests, with error bars representing 1 standard deviation. The red dashed line represents where the test accuracy of the supervised GEKO, unsupervised GEKO or PauliZZ kernel would be equal to the test accuracy of the RBF kernel. 
    }
    \label{fig:acc_std_results}
  \end{figure*}

  The data was then scaled between $[-\pi/2, \pi/2]$, and tested 10 times for each method to determine the prediction test accuracy and entropy of entanglement of each kernel. The results are displayed in Figure \ref{fig:acc_std_results}.

\begin{figure*}
    \centering
    \includegraphics[width=\textwidth]{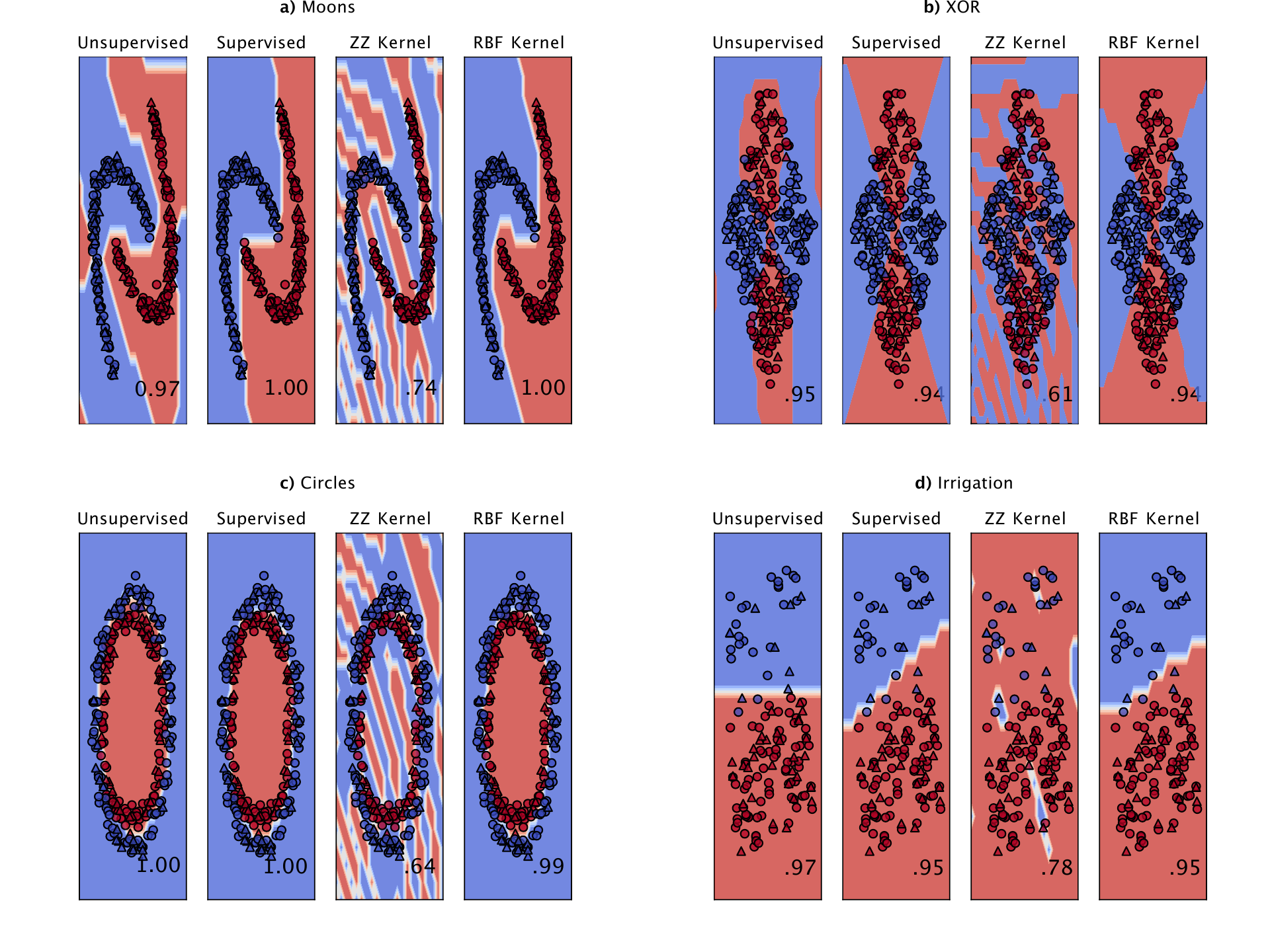}
    \caption[QSVM Decision Boundaries for Datasets]{Decision boundaries for the Moons, XOR, Circles, and Irrigation datasets. Training data and testing data are represented by circles and triangles, respectively. The decision bounds vary from red to blue, with darker shades being higher confidence and white being the region of least confidence.}
    \label{fig:decision_bounds}
  \end{figure*}

  The results show that over the ten test sets for each kernel, both the supervised and unsupervised GEKO kernels consistently outperform the PauliZZ kernel, and perform comparably to the RBF kernel, on each dataset. When viewing the decision boundaries produced by each technique, examples for Moons, XOR, Circles, and Irrigation data are shown in Figure \ref{fig:decision_bounds}, it becomes more clear why certain techniques achieve higher accuracies than others.
  
\begin{figure*}
    \centering
    \includegraphics[width=\textwidth]{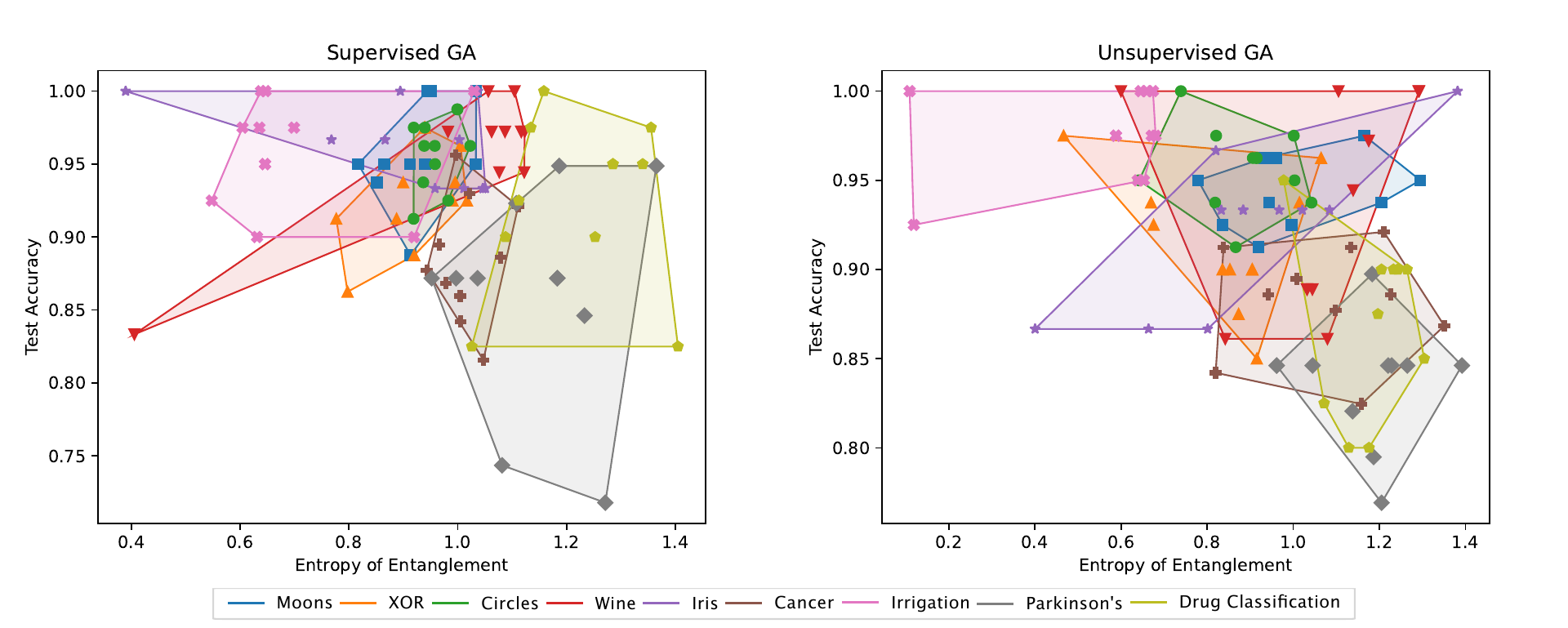}
    \caption[Comparison of the Clustering of Datasets]{Comparison of the clustering of datasets for supervised and unsupervised GEKO kernel entropies compared to test accuracy of the given kernel on Moons, XOR, Circles, Wine, Iris, Cancer, Irrigation, Parkinson's and Drug Classification datasets, coloured blue, orange, green, red, purple, brown, pink, and grey respectively.} 
      \label{fig:ent_trend_res}
  \end{figure*}

  For instance, it can be seen that the GEKO and RBF kernels have smooth, defined decision boundaries, clearly separating the two classes. In contrast, the PauliZZ produces a patchy decision boundary that does not appropriately separate the classes. Over the ten tests, the highest test accuracy GEKO kernel is always equal to or higher than the highest test accuracy RBF kernel, however, there is more variance in the test accuracy achieved. This is not unexpected, as GEKO is a stochastic process, and in real-world applications, the best of several tests would be chosen. In general, the supervised and unsupervised GEKO techniques seem to perform comparably, which implies, that for these datasets, this technique does not require the supervised training of QSVMs as a fitness metric to develop kernels. When viewing the entropy plots, it can be seen that both GEKO methods produce kernels of higher test accuracy than the PauliZZ kernel, with entanglement being fairly independent of the dataset. Further, when linear regression was applied to the test accuracy and entropy data for each dataset, it was seen that the majority had a weakly positive gradient, as seen in Figure \ref{fig:ent_trend_res} (correlation coefficients are given in Table \ref{tab:correlations_table}).
  
  \begin{table*}
    \small
    \centering
    \caption{\label{tab:correlations_table}Correlation of Test Accuracy and Entropy of Entanglement for Each Dataset.}
    \begin{tabular}{@{}lll}
      \hline
      \hline
      Dataset & Supervised Correlation &  Unsupervised Correlation\\
      \hline
      Moons & \textbf{0.202} & \textbf{0.034} \\
      XOR & \textbf{0.228} & -0.072 \\
      Circles & \textbf{0.171} & -0.0366 \\
      Wine & \textbf{0.195} & \textbf{0.064} \\
      Iris & -0.076 & \textbf{0.137} \\
      Cancer & \textbf{0.070} & \textbf{0.002} \\
      Irrigation & \textbf{0.014} & \textbf{0.030} \\
      Parkinson's & \textbf{0.028} & -0.006 \\
      Drug Classification & \textbf{0.039} & -0.037 \\ 
      \hline
      \hline
    \end{tabular}
  \end{table*}
  
  For the supervised method, the Moons, XOR, Circles, Wine, Cancer, Irrigation, Parkinson's and Drug Classification datasets had positive gradients, whereas the Iris dataset had a negative gradient. For the unsupervised method, the Moons, Wine, Cancer, Iris, Cancer, and Irrigation datasets had positive gradients, whereas the XOR, Circles, Parkinson's and Drug Classification datasets had negative gradients. These results imply that for both techniques, there is little evidence to support that as the entropy of the circuit increases, so does the test accuracy, with the evidence being even weaker for the unsupervised technique. These results are somewhat expected, as the unsupervised technique essentially increases the generalisability of the kernel to increase test accuracy, whereas the supervised technique optimises for the specific labelled data. It would be interesting for further research to optimise with a fitness function that combines the two methods, as such creating a model that is highly generalised but also optimised on the specific data. By selecting the optimal kernel from the population based on training fitness, and then only updating the base individual if the validation accuracy is higher, GEKO is forced to maintain generality while increasing accuracy. The results demonstrate that a genetic algorithm can be used to optimise QSVM kernels, producing circuits that outperform manually designed classical and quantum kernels on standard classification tasks.

  \subsection{Quantum Neural Networks}

\begin{figure}
    \centering
    \includegraphics[width=0.5\textwidth]{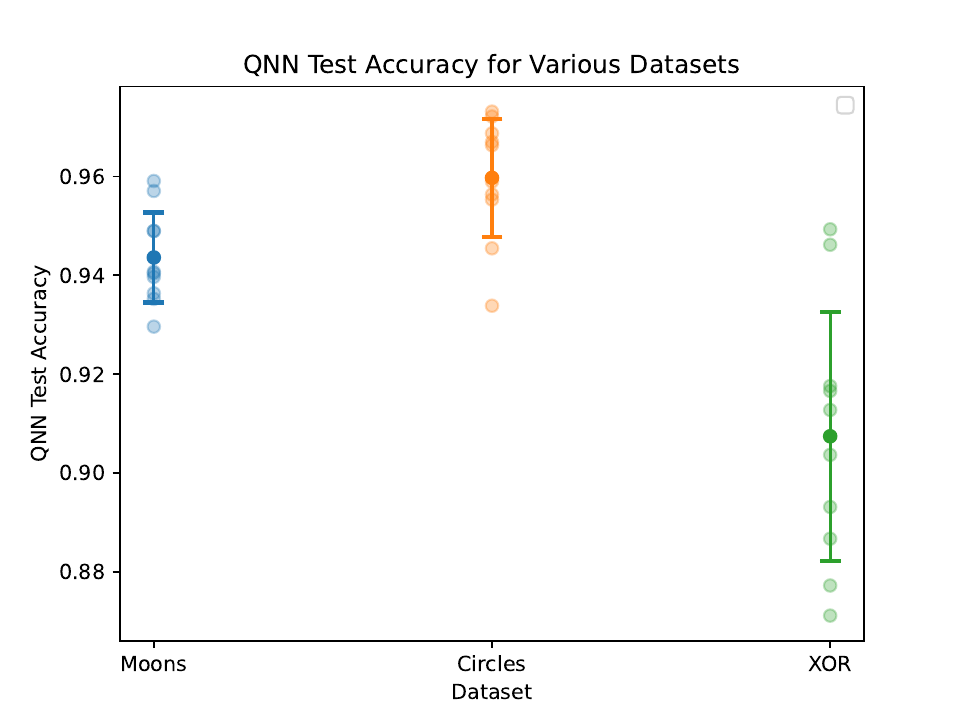}
    \caption[Accuracy of QNN on Various Datasets]{Test accuracy of QNN on Moons, XOR, and Circles datasets. Moons data, Circles data, and XOR data are in blue, orange, and green, respectively. The plots show the comparison of the test accuracy achieved by the QNN over the datasets. The circles represent the average result over the 10 tests, with error bars representing 1 standard deviation, for each dataset.}
    \label{fig:acc_std_results_QNN}
  \end{figure}

  The method was tested on the Moons, XOR, and data sets. Each data set was split into $10\%$ testing data, $80\%$ training data, and $10\%$ validation data. The data was then scaled between $[-\pi/2, \pi/2]$, and tested 10 times for each method to determine the test accuracy. The results are displayed in Figure \ref{fig:acc_std_results_QNN}. 
  
\begin{figure}
    \centering
    \includegraphics[width=0.5\textwidth]{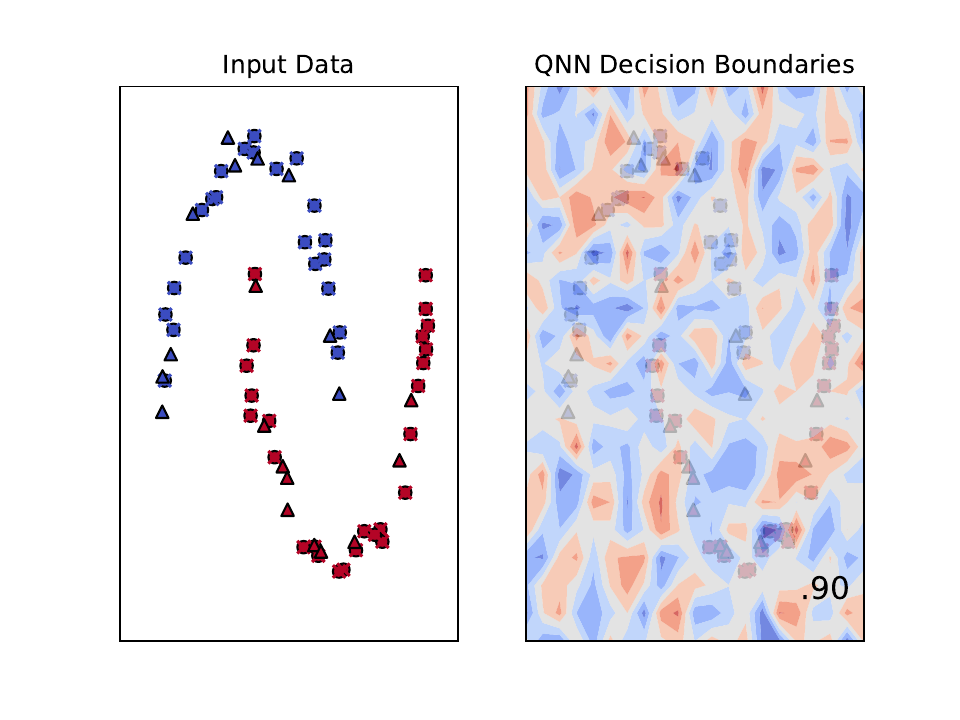}
    \caption[QNN Decision Boundaries for Datasets]{QNN decision boundaries for the Moons, XOR, and Circles datasets. Training data, validation data, and testing data are represented by circles, crosses, and triangles, respectively. The decision bounds vary from red to blue, with darker shades being higher confidence and white being the region of least confidence.}
    \label{fig:decision_bounds_QNN}
  \end{figure}

  The results show that over the three test data sets, the QNN is able to achieve high test accuracy. When viewing the decision boundaries produced by each technique, shown in Figure \ref{fig:decision_bounds_QNN}, it becomes more clear why the QNNs achieve high accuracies; it can be seen that the QNNs have smooth, defined decision boundaries, clearly separating the two classes.
  
\begin{figure}
    \centering
    \includegraphics[width=0.5\textwidth]{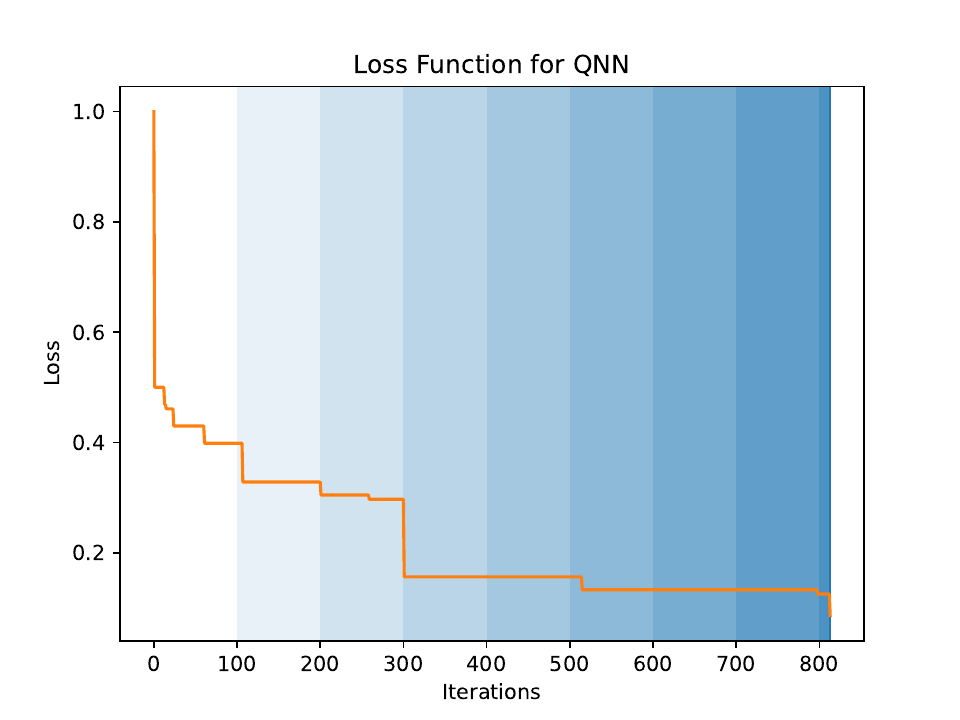}
    \caption[Loss Function for improvements for a QNN]{Loss function for improvements for a 2 qubit QNN. The loss is represented as an orange line, and each time there was an improvement of the QNN feature map is represented by a progressively darker shaded area of blue.}
    \label{ref:2q_9_moons_improvements}
  \end{figure}

  Further, it can be seen that GEKO allows the QNN to overcome plateaus in the optimisation by altering the feature map, an example of this can be seen in Figure \ref{ref:2q_9_moons_improvements}.

  By selecting the optimal feature map from the population based on training fitness, and then only updating the base individual if the validation accuracy is higher, GEKO is forced to maintain generality while increasing accuracy. The results demonstrate that a genetic algorithm can be used to optimise QNN feature maps, producing circuits that produce high test accuracy results.

  \section{Conclusion} \label{sec:conclusion}

  In this work, an approach for optimising QSVM kernels and QNN feature maps using a genetic algorithm, based on GASP \cite{creevey_gasp_2023}, GEKO, was presented. The approach has been shown to outperform manually designed kernels on standard toy datasets, and produced high test accuracy QNNs, demonstrating the potential of this technique for improving the performance of QSVMs and QNNs. The QSVM results suggest that the method may be useful for identifying patterns in complex datasets, especially those that are difficult to analyse with classical machine learning techniques. This efficiency and improved performance make the approach a promising tool for researchers and practitioners in various fields, such as finance, healthcare, and materials science, where data analysis and prediction are crucial. There are several avenues for future research in this area. A potential direction is to investigate the effectiveness of the approach on larger and more complex datasets, such as those in real-world applications. Another interesting area for exploration is the use of other genetic algorithms or optimisation techniques, such as simulated annealing or particle swarm optimisation, to further improve the performance of QSVMs and QNNs. It would be interesting to research to optimise the QSVMs with a fitness function that combines supervised and unsupervised techniques. In conclusion, this study provides an effective approach for optimising QSVM kernel circuits and QNN feature maps using a genetic algorithm. The results suggest that this technique has significant potential for improving the performance of QSVMs and QNNs on standard classification tasks, as well as for identifying patterns in complex data in various fields.

  \section{Acknowledgements} \label{sec:acknowledgements}
    This research was supported by the University of Melbourne through the establishment of the IBM Quantum Network Hub and supported in part by the Australian Research Council Centre of Excellence for Quantum Biotechnology (CE230100021) at the University. FMC was supported by Australian Government Research Training Program Scholarships. This research was supported by The University of Melbourne’s Research Computing Services and the Petascale Campus Initiative. 

    \section{Author contributions statement}
        F.M.C conceived the project with input from J. A. H., M. E. S., and L. C. C. H. The computational framework was created by F.M.C., who also performed the experimental calculations. All authors had input in writing the manuscript.
    
    
    \section{Data availability} \label{sec:data}
        The datasets generated during and/or analysed during the current study are available from the corresponding author on reasonable request.

    \pagebreak

    \clearpage
  \bibliography{main.bib}
\end{document}